\documentclass[12pt]{article}
\usepackage{graphics,color}
\begin{document}
\thispagestyle{empty}
\noindent\
\\
\\
\\
\begin{center}
\large \bf  Composite Weak Bosons at the Large Hadronic Collider
\end{center}
\hfill
 \vspace*{1cm}
\noindent
\begin{center}
{\bf Harald Fritzsch}\\
Department f\"ur Physik\\ 
Ludwig-Maximilians-Universit\"at\\
M\"unchen, Germany \\
\end{center}

\begin{abstract}

In a composite model of the weak bosons the p-wave bosons are studied.
The state with the lowest mass is identified with the boson, which has been discovered at the LHC. Specific properties of the excited bosons are discussed, in particular their decays into weak bosons and photons. Recently a two photon signal has been observed, which might come from the decay of a neutral heavy boson with a mass of about 0.75 TeV. This particle could be an excited weak tensor boson.  

\end{abstract}

\newpage

In the Standard Model of the electroweak interactions the masses of 
the weak bosons and of the leptons and quarks 
are generated by a spontaneous breaking of the electroweak symmetry. 
Besides the weak bosons a scalar boson must exist ( "Higgs boson"). In 2012 a scalar boson has been discovered at the LHC. This boson with the mass of 125 GeV might be the Higgs boson, since it decays into two weak bosons or into two photons (ref.(1,2)).\\

In quantum chromodynamics the masses of the hadrons are due to the field energy of the quarks and gluons inside the hadrons. The hadron masses can be calculated in terms of the QCD scale parameter $\Lambda_c$, if the quark masses are neglected. Thus in the Standard Theory there are two different ways to generate the masses of the particles:\\
a) the dynamical mass generation for the masses of the baryons and mesons,\\
b) the Higgs mechanism for the masses of the weak bosons, the leptons and the quarks.\\ 

More than 99 \% of the proton mass is generated dynamically. The contribution of the Higgs mechanism, which generates the masses of the quarks inside the proton, is less than 1 \%. Perhaps the masses of the weak bosons are also generated dynamically. This is possible, if the weak bosons are not elementary gauge bosons, but bound states. Then there should also exist excited weak bosons, and the scalar boson, observed in the LHC experiments, could be an excited weak boson.\\ 

We assume that the weak bosons consist of a fermion and its antiparticle, which are denoted as "haplons" (see also ref.(3,4,5,6)). The new confining gauge theory is denoted as Quantum Haplodynamics ($QHD$). The $QHD$ mass scale is given by a mass parameter $\Lambda_h$, which determines the size of the weak bosons. The haplons interact with each other through the exchange of massless gauge bosons. The number of these gauge bosons depends on the gauge group, which is unknown. It might be $SU(3)$, as the gauge group of $QCD$.\\ 

Two types of haplons are needed as constituents of the weak bosons, denoted by $\alpha$ and $\beta$. The three weak bosons are bound states of two haplons with the electric charges (+1/2) 
and (-1/2).\\  

In the absence of electromagnetism the weak bosons are degenerate in mass. If the electromagnetic interaction is introduced, the mass of the neutral boson increases due to the mixing with the photon. Details can be found in ref. (7,8).\\

The  $QHD$ mass scale is about thousand times larger than the $QCD$ mass scale (ref.(7,8)). In strong interaction physics above the energy of 1 GeV many resonances exist. We expect similar effects in the electroweak sector. At high energies there should in particular exist excited weak bosons, which will decay mainly into two or three weak bosons. These states can be observed at the $LHC$.\\

The weak bosons consist of pairs of haplons, which are in an s-wave. The spins of the two 
haplons are aligned, as the spins of the quarks in a $\rho$-meson. The first excited states are those, in which the two haplons are in a p-wave. We describe the quantum numbers of these states by $I(J)$. The $SU(2)$-representation is denoted by $I$, $J$ is the total angular momentum.\\ 

There are three $SU(2)$ singlets: $0(0)$, $0(1)$ and $0(2)$, and three $SU(2)$ triplets: $1(0)$, $1(1)$ and $1(2)$. The  singlets consist of a haplon and its antiparticle:
\begin{eqnarray}
 S=& \frac{1}{\sqrt{2}} \left( \overline{\alpha} \alpha +
\overline{\beta} \beta \right) \ .
 \end{eqnarray}
The three triplet states are:
\begin{eqnarray}
T^+ & = &  \overline{\beta} \alpha \; , \nonumber \\
T^- & = & \overline{\alpha} \beta \; , \nonumber \\
T^3 & = & \frac{1}{\sqrt{2}} \left( \overline{\alpha} \alpha -
\overline{\beta} \beta \right) \; .
\end{eqnarray}

We compare these states with the low mass mesons in strong interaction physics, in which the quarks are in a p-wave: the scalar meson $\sigma$, the vector meson $h_1(1170)$   and the tensor meson $f_2(1270)$. These mesons correspond to the singlet states $0(0)$, $0(1)$ and $0(2)$. The isospin triplet mesons, the scalar meson $a_0(980)$, the vector meson $b_1(1235)$ and the tensor meson $a_2(1320)$, correspond to the bosons $T(0)$, $T(1)$ and $T(2)$.\\

What is the mass spectrum of these excited weak bosons? We assume that the boson $S(0)$ is the particle, discovered at CERN (ref. (1,2)). In this case the mass of $S(0)$ is about 125 GeV. In analogy to QCD we expect that the masses of the other p-wave states are in the range 0.2 - 0.8 TeV. The mass of the $S(1)$-boson should be just above 0.3 TeV. The mass of the 
$S(2)$-boson might be between 0.5 and 0.7 TeV. These estimates of the masses are not precise. The mass of the  $S(2)$-boson might as large as 0.75 TeV, as discussed below.\\

The masses of the $SU(2)$-triplet bosons $T$ are expected to be larger than the masses of the $S$-bosons. The mass of the $T(0)$-boson might be about 0.4 TeV, the mass of the $T(1)$-boson just above 0.5 TeV and the mass of the  $T(2)$-boson above 0.6 TeV.\\

The $S(0)$-boson will decay mainly into two charged weak bosons, into two $Z$-bosons, into a photon and a $Z$-boson or into two photons. Since the mass of $S(0)$ is less than twice the mass of the charged weak boson, one of the weak bosons must be virtual.\\
 
The $Z$-boson is a mixture of the state $W^3$ and of the photon. The mixing angle is the weak angle, measured to about 28.7 degrees. Using this angle, we can calculate the branching ratios for the various decays, taking into account the available phase space and assuming that there is no decay into leptons or quarks. The branching ratio for the decay into two charged weak bosons is 70 \%, for the decay into two $Z$-bosons 21 \%, for the decay into a $Z$-boson and a photon 3 \% and for the decay into two photons 2 \%. \\

The $S(0)$-boson might also decay into two or more quarks or leptons. These decays cannot be calculated, but we do not expect that the decay rates into two leptons or quarks are given by the masses of the leptons or quark, as they would be, if the new boson would be the Higgs boson.\\  

The bosons $S(1)$ and $S(2)$ and the nine $T$ - bosons have a much higher mass as the $S(0)$-boson. They will decay mainly into three or four weak bosons or photons, e.g. into two charged weak bosons and a $Z$-boson, into two charged weak bosons and a photon or into four charged weak bosons etc. Decays into two weak bosons, into one weak boson and a photon and into two photons would be suppressed, also the decays into lepton or quark pairs.\\

The bosons $T^+$ and $T^-$ can decay into a charged weak boson and a $Z$-boson or a photon, however the branching ratios for these decays will be very small. The main decay modes of the 
$T^+$ would be:\\

$T^+ \Longrightarrow  (W^+ + Z + Z) $,\\

$T^+ \Longrightarrow  (W^+ + Z + \gamma ) $,\\

$T^+ \Longrightarrow  (W^+ + \gamma + \gamma) $.\\

Recently a two photon signal was observed at the LHC, which might be due to the production of a new boson with a mass of 0.75 TeV, decaying into two photons (ref. (9)). This boson might be the $S(2)$-boson or the neutral $T(2)$-boson - both can decay into two photons. If these two bosons have approximately the same mass, the two photon signal would be enhanced.\\ 

In this case also the decay of the two bosons into a $Z$-boson and a photon should be observed. The $S(2)$-boson or the neutral $T(2)$-boson decay into a $Z$-boson and a photon with a rate, which is six times larger than the rate for the decay into two photons. The $Z$-boson can only be observed, if it decays into a muon-pair. The rate for the decay into a muon pair with the invariant mass of the $Z$-boson and a photon is fifty times less than the rate for the decay into two photons. This decay should soon be observed at the Large Hadronic Collider.\\

\end{document}